\documentclass[aps,prr,superscriptaddress,notitlepage,twocolumn,bibnotes
]{revtex4-1}
\usepackage{graphicx,amsmath,mathtools,amsfonts}
\usepackage{bm}
\usepackage[normalem]{ulem}
\usepackage[usenames, dvipsnames]{xcolor}
\usepackage{multirow}
\usepackage{tabularx}
\usepackage{soul} 
\usepackage{svg}  
\usepackage{hyperref}
\hypersetup{colorlinks=true, linkcolor=blue, citecolor=blue, urlcolor=blue}
\usepackage{comment}

\newcommand\blue[1]{{\color{black}#1}}

\begin{document}

\title{
Preempted phonon-mediated superconductivity in the infinite-layer nickelates
}

\author{Q. N. Meier}
\thanks{Authors with equally important and crucial contributions.}
\affiliation{Univ. Grenoble Alpes, CNRS, Grenoble INP, Institut Néel, 25 Rue des Martyrs, 38042, Grenoble, France}
\author{J. B. de Vaulx}
\thanks{Authors with equally important and crucial contributions.}
\affiliation{Univ. Grenoble Alpes, CNRS, Grenoble INP, Institut Néel, 25 Rue des Martyrs, 38042, Grenoble, France}
\author{F. Bernardini}
\affiliation{Dipartimento di Fisica, Universit\`a di Cagliari, IT-09042 Monserrato, Italy}
\author{A. S. Botana}
\affiliation{Department of Physics, Arizona State University, Tempe, AZ 85287, USA}
\author{X. Blase}
\affiliation{Univ. Grenoble Alpes, CNRS, Grenoble INP, Institut Néel, 25 Rue des Martyrs, 38042, Grenoble, France}
\author{V. Olevano}
\affiliation{Univ. Grenoble Alpes, CNRS, Grenoble INP, Institut Néel, 25 Rue des Martyrs, 38042, Grenoble, France}
\author{A. Cano}
\email{andres.cano@neel.cnrs.fr}
\affiliation{Univ. Grenoble Alpes, CNRS, Grenoble INP, Institut Néel, 25 Rue des Martyrs, 38042, Grenoble, France}

\date{\today}

\begin{abstract}

Nickelate superconductors are outstanding materials with intriguing analogies with the cuprates. These analogies suggest that their superconducting mechanism \blue{may be} unconventional, although this fundamental question is currently under debate. 
Here, we scrutinize the role played by electronic correlations in enhancing the electron-phonon coupling in the infinite-layer nickelates and the extent to which this may promote superconductivity. 
Specifically, we use {\it ab initio} many-body perturbation theory to perform state-of-the-art $GW$ and Eliashberg-theory calculations. 
We find that the electron-phonon coupling is effectively enhanced compared to density-functional-theory calculations.
This enhancement may lead to low-$T_c$ superconductivity in the parent compounds already. However, it remains marginal in the sense that it cannot explain the record $T_c$s obtained with doping.
\blue{
This circumstance implies that conventional superconductivity is} preempted by another \blue{pairing} mechanism in the infinite-layer nickelates.

\end{abstract}

\maketitle

\section{Introduction}
\begin{figure}[b!]
\includegraphics[width=0.475\textwidth]{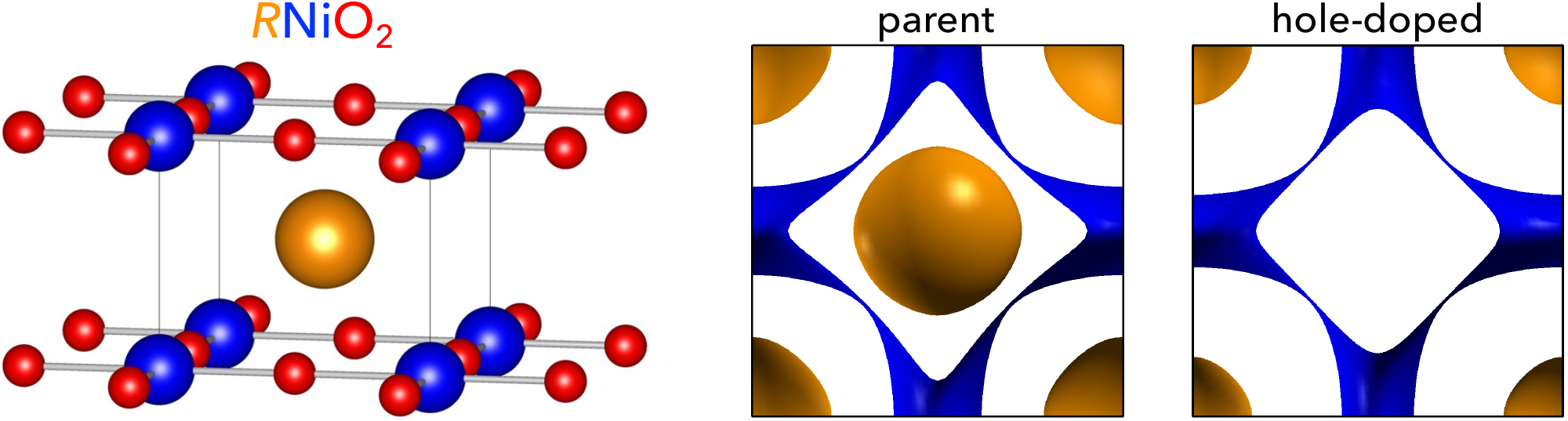}  
\caption{(Left panel) Crystal structure of the infinite-layer nickelates $R$NiO$_2$ with nickel atoms in blue, oxygens in red, and rare-earth in orange. (Middle and right panels) Top view of the characteristic Fermi surface of these systems with the main Ni-3$d_{x^2-y^2}$ sheet in blue and the additional rare-earth derived sheets (self-doping) in orange.
}
    \label{FS}
\end{figure}

After many years of consideration, superconductivity 
in nickel oxides has eventually been discovered \cite{hwang19a,ariando20,hu20,hwang20Pr-a,hwang21-La,ariando22-La,mundy21,sun23}. This finding has sparked a renewed interest in these systems, as they are believed to be unconventional superconductors with intriguing analogies to the high-$T_c$ cuprates \cite{cano21-jetp,tao21-crec,arita22-rpp}. \blue{The first nickelates found to be superconducting are the} infinite-layer nickelates $R$NiO$_2$ ($R=$ rare-earth) \blue{upon hole doping}. 
\blue{These systems} display a layered structure with a square planar coordination of the nickel atoms (see Fig. \ref{FS}) that have the same nominal $3d^9$ filling of the cuprates in their parent phases (Ni$^{1+}$ is isoelectronic with Cu$^{2+}$). In fact, at the density-functional-theory (DFT) level, the Fermi surface of these systems resembles the one of the cuprates and is similarly dominated by 3$d_{x^2-y^2}$ states (see Fig. \ref{FS}) \cite{pickett-prb04,botana20prx,jia20prx}. 
However, \blue{in contrast to the cuprates, there is} an increased out-of-plane mixing of the atomic orbitals resulting in the \blue{so-called} self-doping effect. 
\blue{As a result,} the Fermi surface of the parent compounds displays two additional electron pockets  that are mainly associated with rare-earth derived \blue{5$d$} states. Introducing hole doping as in Nd$_{0.8}$Sr$_{0.2}$NiO$_2$ reduces these pockets and eventually promotes superconductivity with maximal $T_c$ around $20~$K. 

\blue{
In order to understand nickelate superconductivity, one important question that needs to be addressed is whether the Cooper-pairing mechanism in these systems relies on the conventional electron-phonon coupling or not. 
Initial work by Nomura {\it et al.} ruled out phonon-mediated superconductivity from} 
DFT calculations \cite{arita19}, \blue{in line with recent results} by Di Cataldo {\it et al.} for hydrogen-intercalated systems \cite{held23}. \blue{That} conclusion, however, has been revisited by Li and Louie, who \blue{have} used a advanced theoretical framework \cite{louie22}. In particular, they have considered 
the hole-doped compound Nd$_{0.8}$Sr$_{0.2}$NiO$_2$ and included 
electronic correlations by performing {\it ab initio} many-body perturbation theory calculations. 
Many-body effects have been reported to increase the self-doping effect in these systems \cite{cano20b}, which further would enable phonon-mediated superconductivity according to \cite{louie22}. 
Specifically, Li and Louie
reported an unprecedented enhancement of the electron-phonon coupling in Nd$_{0.8}$Sr$_{0.2}$NiO$_2$
further inducing superconductivity with a $T_c$ 
compatible with the experiments. 

\blue{In this work,} we further scrutinize the possibility of phonon-mediated superconductivity in the infinite-layer nickelates. Similarly to \cite{louie22}, we perform {\it ab initio} many-body perturbation theory calculations in the $GW$ approximation to improve the description of their electronic \blue{structure} compared to DFT \cite{Hedin65}. 
That is, we exploit a Green's function formalism to include dynamical correlations and hence better describe the excited states \cite{martin16}. However, in contrast to \cite{louie22}, we treat the dynamical screening of the Coulomb interaction exactly ---via direct numerical integration--- and then \blue{systematically} compare it \blue{with the} results obtained by plasmon-pole models. Thus, from more accurate calculations, we find that the electron-phonon coupling remains too weak to explain the measured $T_c$ in 
\blue{Nd$_{0.8}$Sr$_{0.2}$NiO$_2$.
At the same time, we intriguingly find rather weak but non-negligible superconducting instabilities in the parent compounds.
}

\blue{To shed light} on the subtle interplay between the different electronic-structure features relevant for phonon-mediated superconductivity in these \blue{nickelates, we further analyze the effect} 
of pressure.  
We find that the \blue{trends in the} electron-phonon coupling fail to explain the reported increase of $T_c$ with pressure \cite{wang22-natcomm}, 
essentially because the increase in the self-doping is overtaken by the reduction in the density of states at the Fermi level. 
\blue{As a result, possible electron-phonon-mediated superconductivity appears to be preempted by another mechanism in the infinite-layer nickelates.
}

\section{Methods}\label{sec:Methods}

DFT calculations were performed with the {\sc ABINIT} code using the norm-conserving pseudo-potentials from PseudoDojo with the PBE form of the generalized gradient approximation \cite{ABINIT,dojo,PBE}\blue{, and additionally with the LDA and the hybrid HSE06 exchange-correlation functionals \cite{LDA,HSE06}.} The Nd substitution with Sr was included with the virtual crystal approximation. \blue{The pertinence of this method has been carefully addressed in \cite{louie22}.} 
The calculations were converged with a $12\times12\times14$ Monkhorst-Pack $k$-mesh with a 100~\blue{Ha} cutoff for the wavefunctions and a 0.01 \blue{Ha} smearing. 
For the calculations at different pressures, the lattice parameters were optimized using a convergence threshold of 0.001 Kbar. 

One-shot $G_0W_0$ calculations were performed starting from the DFT electronic structure and using an unshifted 6$\times$6$\times$6 grid $k$-point sampling of the Brillouin zone. The screening was calculated using a wavefunction cutoff of 40 Ha, a dielectric matrix size of 939 plane-waves (15~Ha) to take into account local-field effects, and summing over 170 bands. Convergence on the $GW$ self-energy was obtained with a cutoff on the exchange and correlation parts of 40 and 15 Ha, respectively, and a sum over 200 bands 
\blue{
when using either the contour-deformation (CD) integration technique or the Godby-Needs (GN) plasmon-pole model for the screening.
In the case of the Hybertsen-Louie (HL) plasmon-pole model, however,}
1400 bands were needed to achieve convergence 
(\blue{the slow convergence of this model with respect to the number of bands has previously been pointed out in \cite{Riganese2011})}. The CD integration was performed using 30 frequencies on a logarithmic mesh along the imaginary axis, and 200 on the real axis to evaluate the pole residuals. $GW$ quasiparticle corrections were calculated for 29 bands around the Fermi level \blue{starting from the bottom of the O-$2p$ manifold} 
by solving the quasiparticle equation. 
Both the DFT eigenvalues and the $GW$ energies were interpolated with Wannier90 \cite{Wannier90}. Specifically, we used 17 Wannier functions associated with the O-2$p$, Ni-3$d$, and $R$-5$d$ states and one additional interstitial-$s$ state as in \cite{louie22}. \blue{For the the La compound we used 7 additional Wannier functions associated with La-$4f$ states}. 

Phonon calculations were performed using the \textsc{Quantum Espresso} package \cite{QE,Giannozzi_2017}. We used a $k$-grid of $18\times 18 \times 18$ with a planewave cutoff of 125~Ry for the corresponding DFT calculations. 
\blue{
In these calculations, we stick to PBE for the exchange-correlation functional as this is a reliable choice for the structural properties of interest (see e.g. \cite{cano22-prm}).} Phonon spectra and displacement potentials were calculated on a $q$-point grid of $6\times 6\times 6$  using density functional perturbation theory.

For the calculations of electron-phonon coupling matrices and Eliashberg equations we use the wannier-interpolation method as implemented in \textsc{EPW} package \cite{Giustino2007,Ponce2016,Margine2013}. Specifically, we used the $GW$ electronic structure combined with the phonon frequencies and electron-phonon matrix elements obtained from DFT. A coarse grid of $6\times 6 \times 6$ was used for both the electronic structure and the phonons, with interpolated fine grids of $36\times 36\times 36$ and $18 \times 18 \times 18$ for the electronic and phononic parts respectively. The presented Eliashberg spectral function is computed as:
\begin{align}
\alpha^2F(\omega)
=
\dfrac{1}{2 N_F}
&\sum_{
nm,\nu}\int_\text{BZ} {d\mathbf{k}\over (2\pi)^3}\, {d\mathbf{q}\over (2\pi)^3}
\left|g_{mn,\nu}(\mathbf{k},\mathbf{q})\right|^2
\nonumber \\
&\times \delta(\varepsilon_{n\mathbf{k}})
\delta(\varepsilon_{m\mathbf{k+q}})\delta (\omega-\omega_{\nu \mathbf{q}}),
\end{align} 
where $N_F$ is the density of states at the Fermi level, $g_{mn,\nu}(\mathbf{k},\mathbf{q})$ represents the electron-phonon matrix elements, $\varepsilon_{n\mathbf{k}}$ is the (quasi-)particle energy with respect to the Fermi level and $\omega_{\nu \mathbf{q}}$ is the phonon frequency.
Further, the electron-phonon coupling matrix elements are calculated on the Fermi surface as
\begin{align}
 \lambda_{\mathbf{k}\mathbf{k'}}  = \dfrac{1}{ N_F}&\sum_{
nm,\nu}\int_0^{\infty} d\omega 
\dfrac{\left|g_{mn,\nu}(\mathbf{k},\mathbf{k-k'})\right|^2}{\omega}
\nonumber \\
&\times \delta(\varepsilon_{n\mathbf{k}})
\delta(\varepsilon_{m\mathbf{k'}}) \delta (\omega-\omega_{\nu \mathbf{k-k'}}).
\end{align}
In these calculations, we considered an energy window of 0.5 eV centered at the Fermi level, with an electronic smearing of 0.05 eV and phonon smearing of 0.5 meV.

\section{Results}

\subsection{
Hole-doped Nd$_{0.8}$Sr$_{0.2}$NiO$_2$ \label{sec:electronic_structure} 
}
\blue{
We start our investigation on Sr-doped NdNiO$_2$ which, experimentally, has been shown to reach its maximal superconducting $T_c$ near $20\%$ doping. In the following, we present our findings of the electronic structure and electron-phonon coupling in Nd$_{0.8}$Sr$_{0.2}$NiO$_2$. 
}

\subsubsection{Electronic structure}

\begin{figure*}[t!]
\includegraphics[width=0.975\textwidth]{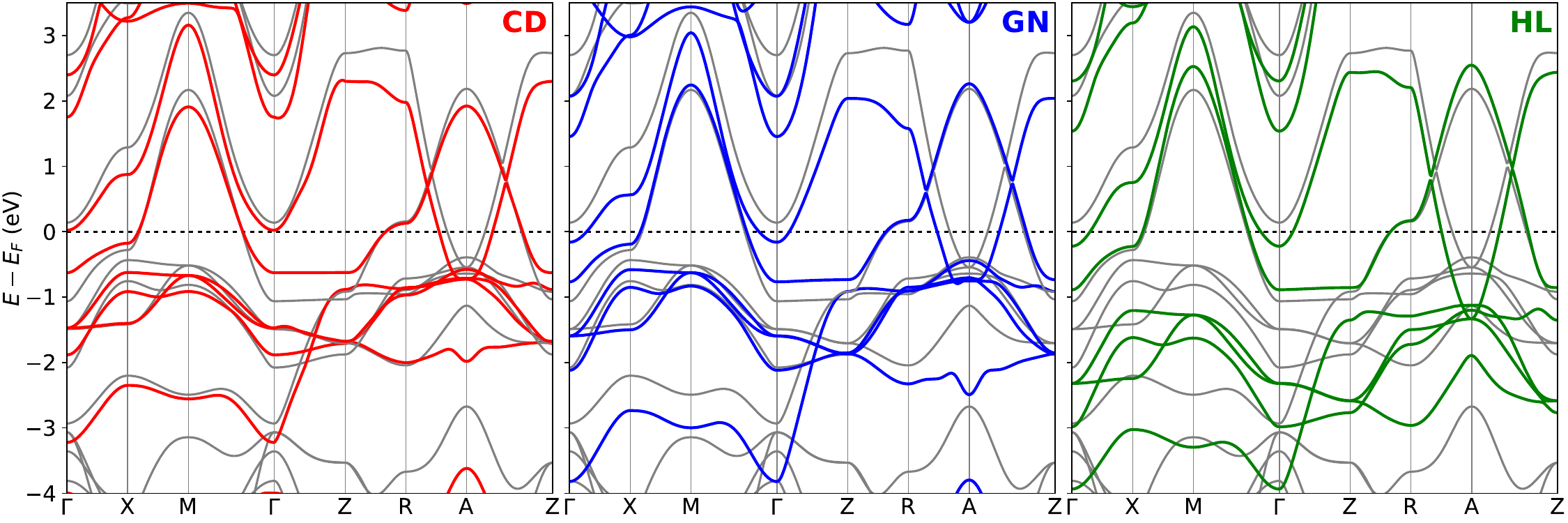}    
    \caption{
    Band structure of Nd$_{0.8}$Sr$_{0.2}$NiO$_2$ near the Fermi level calculated within the $GW$ approximation for the self-energy.  
    The different panels correspond to different treatments of the screening via the contour deformation technique (red), the Godby-Needs plasmon-pole model (blue) and the Hybertsen-Louie one (green), with the DFT result in gray. Compared with the numerically exact $GW$-CD method, the plasmon-pole models tend to overestimate the self-doping effect ({\it i.e.} the dipping of the bands down the Fermi level at $\Gamma$ and A). 
    }
    \label{GW-comparison}
\end{figure*}
\blue{
First, we consider the electronic structure of Nd$_{0.8}$Sr$_{0.2}$NiO$_{2}$.} Figure \ref{GW-comparison} shows the computed band structure of 
\blue{this system} at the \blue{DFT and} many-body $GW$ level for different treatments of the dynamical correlations (see also Fig. \ref{GW-comparison.all}). \blue{At the DFT level, the self-doping pocket at $\Gamma$ vanishes, while the self-doping at the $A$ point remains strong. However, from the $GW$ results,} we confirm a general enhancement of the self-doping due to electronic correlations \blue{which tends to partially restore the electron pocket at $\Gamma$. This is accompanied by a decrease in the bandwidth of the bands crossing the Fermi level, which further translates into a significant increase in the density of states (DOS) at the Fermi level (see Fig. \ref{GW-comparison.all}).} 
At the same time, we find differences \blue{in these changes} depending on how the screened Coulomb interaction is described in practice. Specifically, Fig. \ref{GW-comparison} compares calculations employing numerically exact contour-deformation (CD) integration technique as well as the popular Godby-Needs (GN) and Hybertsen-Louie (HL) plasmon-pole models \cite{GNref,HLref}. $GW$-CD yields a low-energy electronic structure that is \blue{the most}
similar to DFT, \blue{while both GN and HL plasmon-pole models seem to slightly overestimate the many-body corrections and the self-doping effect with GN being in better overall agreement with the CD result in this case (see also Fig. \ref{GW-comparison.all}) \cite{note}.
}

\subsubsection{Electron-phonon coupling}

\begin{figure}[t!]
    \centering
    \includegraphics[width=0.45\textwidth]{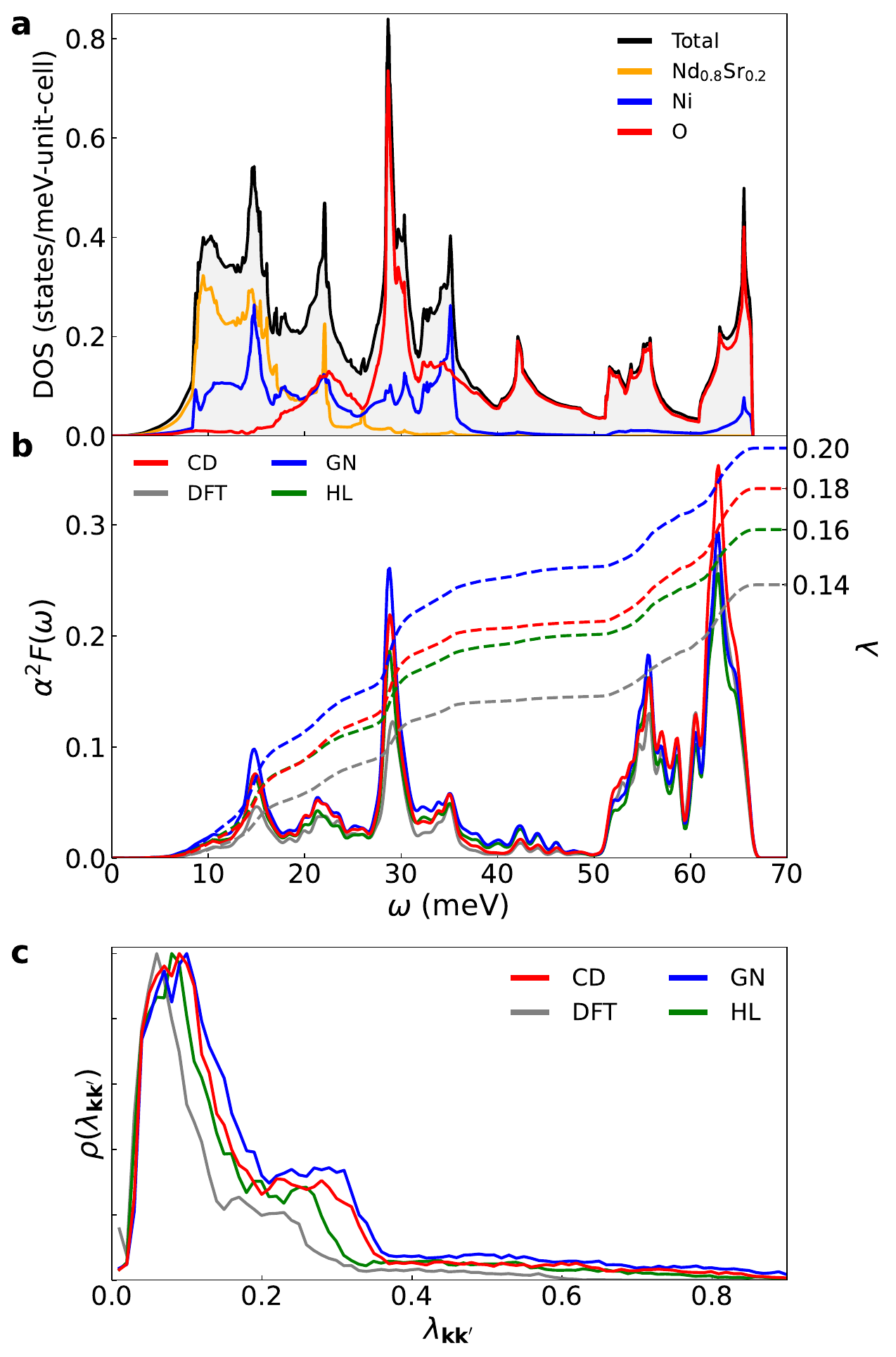}
    \caption{(a) Total and site-resolved phonon density of states, (b) Eliashberg spectral function $\alpha^2F(\omega)$ (solid lines) and corresponding cumulative coupling constant $\lambda (\omega) = \int_0^\omega d\omega'\alpha^2F(\omega')/\omega'$ (dashed lines) and (c) distribution of electron-phonon coupling strength $\lambda_{\mathbf{k}\mathbf{k}'}$ calculated for Nd$_{0.8}$Sr$_{0.2}$NiO$_2$. The different curves in (b) and (c) correspond to different approximations for the electronic structure [DFT vs $GW$ using the numerically exact contour-deformation (CD) technique as well as GN and HL plasmon-pole models].}
    \label{fig:a2f}
\end{figure}

Next, we address the influence of correlations on the electron-phonon coupling and its implications for superconductivity \blue{in Nd$_{0.8}$Sr$_{0.2}$NiO$_2$. To do this, we start by calculating the phonon DOS and phonon spectrum, which are shown in Fig. \ref{fig:a2f} (a) and \ref{phonons-comparison} respectively. We find that}
the lightest O atoms dominate the phonon DOS above 40~meV and produce the main feature at 30~meV. 
The Ni and Nd$_{0.8}$Sr$_{0.2}$ related modes are softer and contribute to the phonon DOS only below 40~meV and 30~meV respectively. Specifically, the 35~meV feature is due to mixed Ni-O contributions, the 22.5~meV due to Nd$_{0.8}$Sr$_{0.2}$-Ni-O ones, while the 15~meV and 10~meV ones are due to Nd$_{0.8}$Sr$_{0.2}$-Ni instead. We note that, compared with the parent NdNiO$_2$ compound, the O modes soften while the Ni ones harden by $\sim 4$ meV (see Fig. \ref{phonons-comparison}).

\blue{
In order to quantify the corresponding electron-phonon coupling and to identify possible phonon-mediated superconducting instabilities, we calculate the Eliashberg spectral function $\alpha^2F(\omega)$ and the total electron-phonon coupling constant $\lambda= 2\int d\omega\alpha^2F(\omega)/\omega$ (see Sec. \ref{sec:Methods}). 
These quantities are shown in Figure \ref{fig:a2f} (b) for Nd$_{0.8}$Sr$_{0.2}$NiO$_2$. 
From the comparison with the phonon DOS in Fig. \ref{fig:a2f} (a), we find the main contributions to $\lambda$ mainly come from the hard O modes with a subdominant contribution from the softer Ni modes. 
In general, the renormalization of the quasi-particle energies due to correlations tends to enhance the electron-phonon coupling compared to DFT values (see e.g. \cite{faber2011,yin2013, Li2019}).
This enhancement, however, turns out to be marginal in Nd$_{0.8}$Sr$_{0.2}$NiO$_2$ with $\lambda$ increasing from 0.14 to just 0.18 in the numerically exact $GW$-CD case, while the plasmon-pole models yield 0.16 (HL) and 0.2 (GN) \cite{note}. These values are summarized in Table \ref{t:Tc}. 
}

\blue{
From $\lambda $, an estimate of the critical temperature for phonon-mediated superconductivity can be obtained according to the McMillan-Allen-Dynes formula $
T_c = {\hbar \omega_{\rm log} \over 1.2 k_B}
\exp\big(
{-1.04(1+\lambda) \over \lambda -\mu^* (1 +0.62\lambda)}\big)$
\cite{mcmillan,allen-dynes}. In this formula, the characteristic phonon frequency 
$\omega_\text{log}$ is obtained from the Eliashberg spectral function as $\omega_\text{log} =\exp\big(\tfrac{2}{\lambda}\int d\omega \tfrac{\log\omega}{\omega}\alpha^2F(\omega)\big)$ while quantity $\mu^*$ is the so-called Coulomb pseudopotential (whose value typically ranges between 0.04 and 0.16). 
The resulting $T_c$'s are summarized in Table \ref{t:Tc}.
Even for a Coulomb pseudopotential as small as $\mu^*=0.05$, the correlation-enhanced electron-phonon coupling seems to be hardly compatible with the experimentally reported superconducting $T_c \sim 20$~K of Nd$_{0.8}$Sr$_{0.2}$NiO$_2$. 
}

The McMillan-Allen-Dynes considerations, however, miss the possibility of having superconductivity with different energy gaps in different parts of the Fermi surface. Consequently, \blue{the calculations described above} may underestimate the corresponding $T_c$ if such an anisotropy becomes relevant. Given the distinct multiband features of infinite-layer nickelates (see Fig. \ref{GW-comparison}), \blue{then} it is more appropriate to analyze the possible emergence of superconductivity in these systems in terms of the full Eliashberg theory. 
To this end, we first calculate the distribution of coupling strength $\lambda_{\mathbf{k}\mathbf{k}'}$ over the Fermi surface.
The $GW$ results are compared with DFT calculations in Figure \ref{fig:a2f} (c). 
This comparison confirms that there is a slight overall enhancement of the electron-phonon coupling due to correlations.
Further, the $GW$ distributions feature a main peak at $\sim 0.1$, a second peak at $\sim 0.27$, and then a broader feature extending up to 1. 
This suggests that the gap function can be different in different parts of the Fermi surface, so that the actual $T_c$ may be higher \blue{than that obtained from the} McMillan-Allen-Dynes formula. 
This is confirmed from the solution of the anisotropic Eliashberg equations from which, in the $GW$-GN case, we obtain $T_c \sim 1$~K (and hence an increase of the effective $\lambda$ from 0.2 to 0.3 due to the multiband features of the system). This $T_c$, however, still fails to explain the experimentally measured one.  

\begin{table}[t!]
\begin{tabular}{l c c c c c }
\hline \hline 
 & \multicolumn{2}{c}{Method} & $\lambda$ & $\omega_{\text{log}}$ (meV) \ & $T_c$~(K) \\
\hline 
\multirow{4}{*}{
Nd$_{0.8}$Sr$_{0.2}$NiO$_2$ 
\quad}
&\multicolumn{2}{c}{DFT}  & 0.136 & 32.3 &  0.00 \\
& $GW$ & CD & 0.183 & 30.5 &  0.02 \\
&      & GN & 0.204 & 28.1 &  0.05 \\
&      & HL  & 0.163 & 29.0 & 0.00  \\
\hline
\multirow{4}{*}{
NdNiO$_2$
}
&  \multicolumn{2}{c}{DFT} & 0.181 & 27.7 & 0.02\\
& $GW$& CD & 0.270 &25.8 &  0.48 \\
&     & GN  & 0.344 & 23.8 & 1.67 \\
&     & HL & 0.250 & 24.3 & 0.27 \\
\hline 
\multirow{3}{*}{
LaNiO$_2$
}
& \multicolumn{2}{c}{DFT} & 0.180 & 28.0 &0.01\\
&  $GW$ & CD & 0.221 & 28.0 & 0.12\\
&       & GN & 0.244 & 26.0 & 0.25\\
\hline  \hline   
\end{tabular}
\caption{Electron-phonon coupling constants $\lambda$ and $\omega_\text{log}$ computed
within DFT and $GW$ and corresponding superconducting transition temperature $T_c$ according to the McMillan-Allen-Dynes formula 
with $\mu^* = 0.05$ \cite{note}.
\label{t:summary-lambda}
} 
\label{t:Tc}
\end{table}

\begin{figure*}[t!]
\centering
\hspace{1.5em}\includegraphics[width=.85\textwidth]{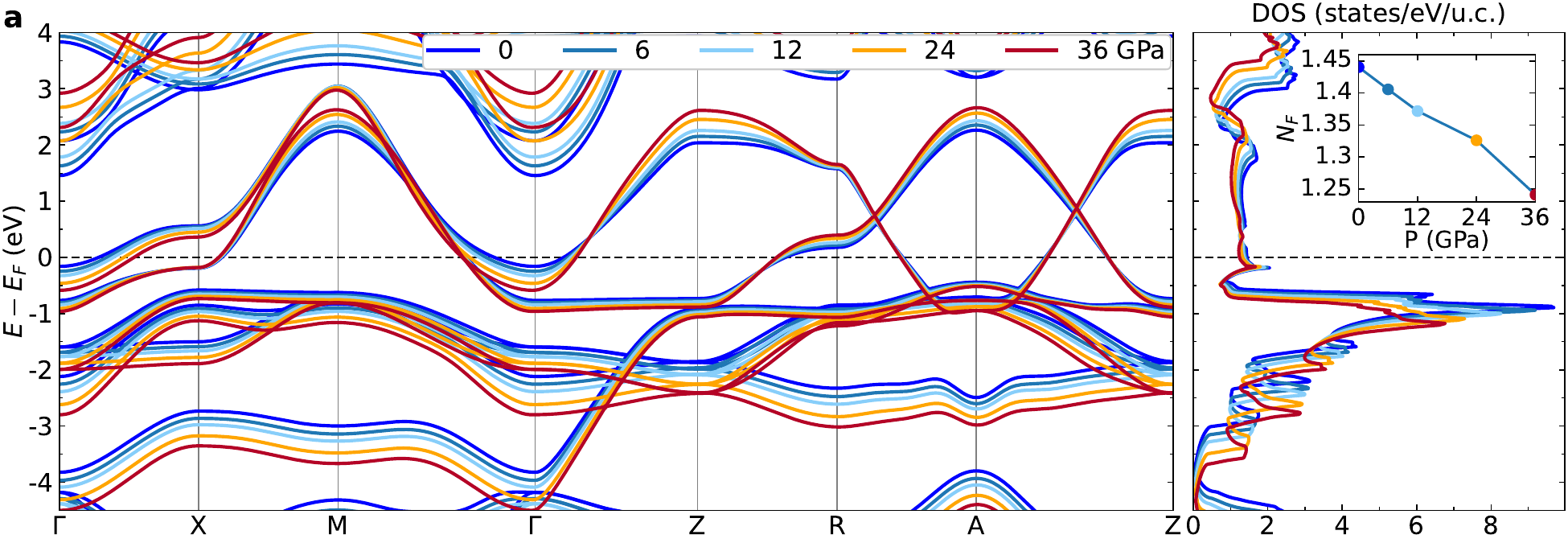}
\includegraphics[width=.9\textwidth]{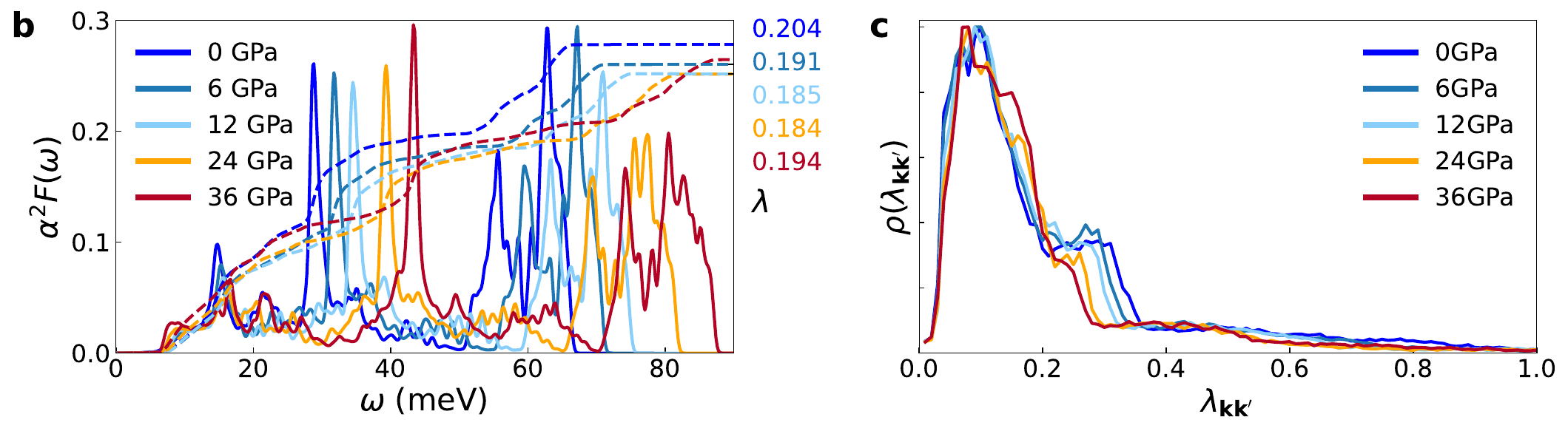}
\caption{
(a) Computed $GW$ band structure and DOS of Nd$_{0.8}$Sr$_{0.2}$NiO$_2$ as a function of pressure. The DOS at the Fermi level is $N_F=$ 1.44, 1.40, 1.37, 1.33 and 1.24 states/eV-unit-cell for 0, 6, 12, 24 and 36 GPa respectively. 
(b) Eliashberg function $\alpha^2F(\omega)$ and electron-phonon coupling constant $\lambda$ and (c) distribution of electron-phonon coupling strength $\lambda_{\mathbf{k}\mathbf{k}'}$ associated with the electronic structures in (a).}
    \label{fig:pressure}
\end{figure*}

\subsection{Parent compounds}

\blue{
From the previous analysis of the effect of correlations on the electron-phonon coupling in Nd$_{0.8}$Sr$_{0.2}$NiO$_2$, we can conclude that the electron pockets at $\Gamma$ and A both provide additional states relevant for electron-phonon interactions. The amount of self-doping then effectively plays an important role with the size of these electron pockets being an indicator of the eventual coupling strength. 
To further investigate this point, we shift our focus to the parent compounds NdNiO$_2$ and LaNiO$_2$ where the self-doping effect is more pronounced. 
We find that, compared to Nd$_{0.8}$Sr$_{0.2}$NiO$_2$, the recovery of the electron pocket at $\Gamma$ in NdNiO$_2$ leads to the increase of $\lambda$ from 0.14 to 0.18 at the DFT level already. 
Then, as illustrated in Fig. \ref{Nd-GW}, the incorporation of further electronic correlations at the $GW$ level increases the eventual amount of self-doping. This translates into the additional increase of
}
the electron-phonon coupling constant $\lambda$ from 0.18 to 0.22 and 0.27 in LaNiO$_2$ and NdNiO$_2$ respectively (see Table \ref{t:summary-lambda} and Fig. \ref{fig:sc_doped_undoped}). 
\blue{This increase} further gives a non-zero superconducting $T_c$ of 0.1~K and
0.5~K according to McMillan-Allen-Dynes formula assuming a Coulomb pseudopotential $\mu^*=$~ 0.05. 
When it comes to the distribution of electron-phonon coupling strength $\lambda_{\mathbf{kk}'}$, we find similar features plus a slight broadening compared to Nd$_{0.8}$Sr$_{0.2}$NiO$_2$ (see Fig.~\ref{fig:sc_doped_undoped}). Thus, the effective effective $\lambda$ associated with the solution of the full Eliashberg equations is expected to undergo a similar increase compared to the isotropic one.
Assuming the same relative change (i.e. from \{0.22, 0.27\} to \{0.33, 0.405\}) the resulting $T_c$ can be estimated to be $T_c \simeq 1.5$~K\,-\,$3.5$~K for $\mu^*=0.05$.
With a larger value of the Coulomb pseudopotential $\mu^*=0.15$, however, the estimated $T_c$ drops to $\lesssim$~0.3~K.

\subsection{Pressure}

\blue{T}he previous analysis confirms that one of the underlying factors that could enable electron-phonon superconductivity in the infinite-layer nickelates manifests through the enhancement of the self-doping effect. 
Such a self-doping traces back to the atomic-orbital overlaps along the out-of-plane direction so that, in principle, it can be tuned by means of applied pressure. \blue{To explore this possibility}, we performed additional calculations as a function of pressure. 

\blue{We first consider Nd$_{0.8}$Sr$_{0.2}$NiO$_2$.} \blue{As the GN plasmon-pole model} has proven to be an accurate approximation to the numerically exact CD method \blue{we stick to that model here.} 
Fig. \ref{fig:pressure} illustrates the changes obtained in the electronic structure of Nd$_{0.8}$Sr$_{0.2}$NiO$_2$ due to hydrostatic pressure. There is an overall increase of the bandwidths so that the DOS at the Fermi level $N_F$ is reduced accordingly. In addition, as expected, the 3D character of the Fermi surface is enhanced as can be seen from the asymmetry of the Ni-3$d_{x^2-y^2}$ band along the $\Gamma$-X-M-$\Gamma$ path vs the Z-R-A-Z one. 
Further, while the self-doping of the Fermi surface remains practically unchanged at A, it increases at $\Gamma$ with pressure. 

When it comes to the phonons, we find a general hardening of all phonon modes $\lesssim 5$~meV/GPa on average (see Figure \ref{phonons-comparison}). Assuming that the coupling constant $\lambda$ remains unchanged, this would imply an increase of $T_c$ with a slope of just $\lesssim 0.03$~K/GPa according to McMillan-Allen-Dynes formula. 
At the same time, we observe a reduction of the DOS at the Fermi level that should reduce the coupling constant since $\lambda = V_\text{e-ph}N_F$ [see inset in Fig \ref{fig:pressure}(a)]. 
In reality, $N_F$ is reduced by $\sim 5\%$ while the calculated $\lambda$ is further lowered by $\sim 10\%$ between 0 and 24 GPa [see Fig. \ref{fig:pressure}(b)]. 
Consequently, the overall change in $\lambda $ is also due to a decrease in the coupling itself. 
This is confirmed in the computed distribution of the matrix elements $\lambda_{\mathbf{k}\mathbf{k}'}$, which shows a small but visible shift towards lower values [see Fig \ref{fig:pressure}(c)]. 
We further obtain a saturation and then a slight increase in the electron-phonon coupling constant for pressures above 24 GPa. However, if the superconductivity was phonon mediated, then the superconducting $T_c$ should initially decrease (rather than increase \cite{wang22-natcomm}) under the application of pressure.

\blue{
Finally, we consider the parent compounds under pressure. The results are illustrated in Fig. \ref{fig:Nd-GW-bands-pressure}. 
Similarly to the previous hole-doped case, the DOS at the Fermi level is reduced.} 
As a result, we find a decrease of the electron-phonon coupling constant $\lambda$ as well as a narrowing of the distribution of electron-phonon coupling strength $\lambda_{\mathbf {kk}'}$ (see Fig. \ref{fig:nd_pressure}).  
The phonon-mediated superconducting instability obtained for the parent compounds is therefore expected to be suppressed by pressure too. 

\subsection{\blue{Robustness of the method}}

\blue{
In the previous sections, we have shown that our results are robust with respect to one important aspect of the $GW$ methodology that is the use of plasmon-pole models vs direct numerical integration for the screening of the Coulomb interaction.
These results, however, may still depend on the initial choice of the DFT functional.
In this section, we analyze this possible
dependence by focusing on the parent compound LaNiO$_2$. This choice is representative from the {\it ab initio} perspective since it avoids limitations related to the modeling of doping as well as to the treatment of the 4$f$ states. 

We first analyze the influence of the DFT-functional choice on the electronic structure itself. 
To consider different starting points for $GW$, we performed electronic-structure calculations using LDA and the hybrid HSE06 exchange-correlation functionals in addition to the PBE functional considered so far. 
In all these calculations, we considered the same PBE lattice parameters to single out purely electronic effects. 
The results are illustrated in Fig. \ref{fig:starting-dft}. 
This figure shows that the initial DFT result can indeed be quite different. 
However, within the $GW$ accuracy, the subsequent $GW$ calculation tends to converge towards a common electronic structure and thereby correct possible drawbacks of the initial assumption. 
This is in fact quite spectacular, in the sense that the Ni-3$d$ derived states are renormalized differently compared with the O-2$p$ ones depending on initial DFT functional: with LDA and PBE the Ni-3$d$ vs O-2$p$ manifolds undergo modest vs substantial shifts respectively, while with HSE06 it is the other way around (see Fig. \ref{fig:starting-dft} and also \cite{cano20b} for additional $GW$@LDA results).  
Also, while performing self-consistent $GW$ calculations is out of the scope of this work, the above observation suggests that the one-shot $GW$ results should be quite comparable to those obtained after reaching self-consistency in the energies. 

The consistency between these results ---again within the $GW$ accuracy--- is also remarkably obtained near the Fermi level, which is the most important region for superconductivity. To further quantify this, we recalculated the electron-phonon coupling constant $\lambda$ combining all the above $GW$ results with the PBE phonons. The result is illustrated in Fig. \ref{fig:lambda-starting}.
This gives a direct measure of the possible spread with respect to the initial DFT functional used for the subsequent $GW$ calculations, from which we obtain $\lambda = 0.24 \pm 0.04$.

The choice of DFT functional also impacts the calculated phonons and electron-phonon matrix elements, and this impact may propagate to the final electron-phonon coupling. We scrutinize this possibility by comparing the results obtained with LDA and PBE phonons (LDA mostly underestimates the $a$ lattice parameter while PBE overestimates $c$). 
Specifically, we performed additional calculations in which the LDA phonons are calculated with the lattice parameters optimized with both LDA and PBE functionals. 
The resulting phonons are quite different, with frequency changes of $\sim$~5~meV or even more. 
However, when it comes to the electron-phonon coupling constant, these differences are surprisingly washed out and the above values are recovered as illustrated in \ref{fig:lambda-starting-phonons}. 
In conclusion, we find that our results are remarkably robust with respect to the choice of the DFT functional.   
}

\section{Discussion}

Our results confirm that, compared to DFT calculations, the electron-phonon coupling in the infinite-layer nickelates is enhanced due to correlations \blue{as included in the $GW$ approximation}. 
\blue{
We have shown that this enhancement is robust with respect to important aspects of the $GW$ methodology, such as the initial approximation for the electronic structure and the subsequent treatment of the screening. 
Our procedure captures the renormalization of the quasiparticle energies only. In doing so, the computed enhancement yields superconducting instabilities in these systems.
However, 
}
the estimated $T_c$ in the hole-doped nickelate Nd$_{0.8}$Sr$_{0.2}$NiO$_2$ remains \blue{way} too \blue{small} compared with the experimental one. 
Consequently, another mechanism seems to take over phonon-mediated superconductivity in this case. 
 
\blue{
Interestingly, we find that the above enhancement results in phonon-mediated superconducting instabilities in the parent NdNiO$_2$ and LaNiO$_2$ compounds also. 
Experimental data suggests more insulating behavior in these systems compared to their doped counterparts even if, in the end, they all remain metallic \cite{hwang21prr}. 
In terms of the Eliashberg theory for phonon-mediated superconductivity, this circumstance would naturally translate into larger values of the Coulomb pseudopotential $\mu^*$ (due to a weakened screening) and hence to a lower, yet non-zero $T_c$ $\lesssim 0.3$~K. Otherwise, the estimated $T_c$ can be as high as $\sim$~3.5~K.}
Experimental evidence of superconductivity has indeed been reported for LaNiO$_2$ \cite{hwang21-La} but, to the best of our knowledge, not for NdNiO$_2$.  
The latter may \blue{simply be due to} sample quality issues\blue{, as it was initially the case for LaNiO$_2$ \cite{hwang19a}, or due to the emergence of competing orders (associated with magnetic or charge-wave instabilities). 
In any case, this point deserves further attention according to our results.}

\section{Conclusions}

We have investigated the possibility of phonon-mediated superconductivity via the correlation enhancement of the electron-phonon coupling in the infinite-layer nickelates. 
This possibility traces back to the distinct self-doping effect that distinguishes these systems from the cuprates.
We have shown that such an enhancement produces superconducting instabilities in the infinite-layer nickelates. This includes the parent compounds where\blue{, in the absence of competing orders, that} type of superconductivity may thus be realized. 
In the hole-doped case, however, the calculated $T_c$ is much lower than the experimentally reported (and gets further reduced with pressure) so that electron-phonon-mediated superconductivity is visibly preempted by another unconventional mechanism in these systems.

\section*{Acknowledgements}

We acknowledge HPC resources from GENCI Grant 2022-AD010913948 and the LANEF Chair of Excellence program for funding.

\bibliography{bib.bib}

\clearpage

\onecolumngrid

\renewcommand{\thefigure}{S\arabic{figure}}
\renewcommand{\thetable}{S\arabic{table}}
\setcounter{page}{1}
\setcounter{figure}{0}
\setcounter{table}{0}

\section*{Supplementary Information}

\begin{center}
    
{\large \bf
Preempted phonon-mediated superconductivity in the infinite-layer nickelates
}
\vspace{1em}
\\
{Q. N. Meier,$^1$ J. B. de Vaulx,$^1$ F. Bernardini,$^2$ 
A. S. Botana,$^3$ X. Blase,$^1$ V. Olevano,$^1$ and A. Cano$^1$}
\\
\vspace{1ex}
{\small \it 
$^1$Univ. Grenoble Alpes, CNRS, Grenoble INP, Institut Néel, 25 Rue des Martyrs, 38042, Grenoble, France\\
$^2$Dipartimento di Fisica, Universit\`a di Cagliari, IT-09042 Monserrato, Italy
\\
$^3$Department of Physics, Arizona State University, Tempe, AZ 85287, USA
}
\end{center}


\begin{figure}[h!]
\includegraphics[width=0.75\textwidth]{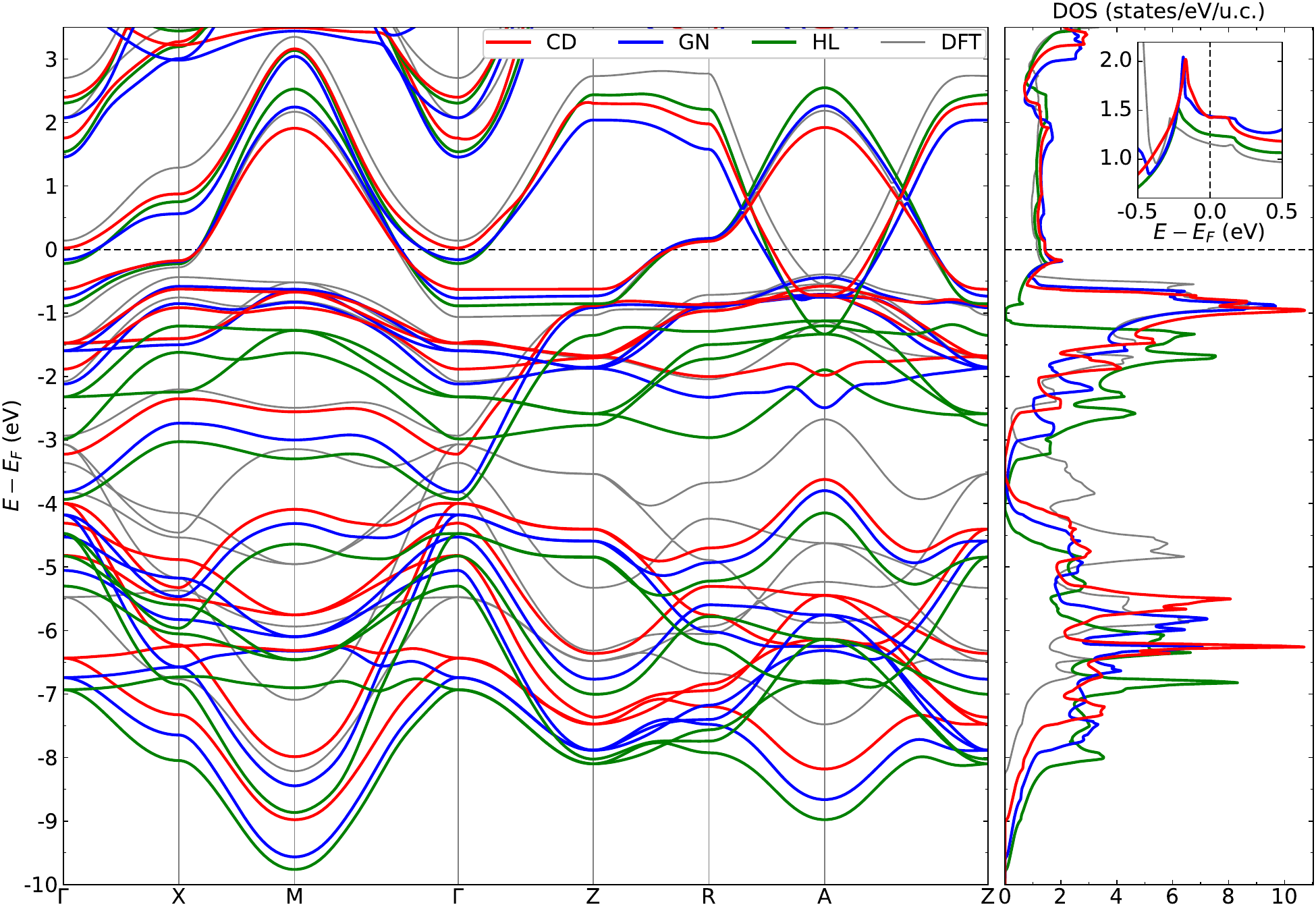}    
    \caption{
    Band structure and density of states (DOS) of Nd$_{0.8}$Sr$_{0.2}$NiO$_2$ calculated within DFT (gray) and $GW$ for different treatments of the screening of the Coulomb interaction (contour-deformation in red, Godby-Needs in blue and Hybertsen-Louie green). 
    }
    \label{GW-comparison.all}
\end{figure}

\begin{figure}[h!]
    \centering
\includegraphics[width=.9\textwidth]{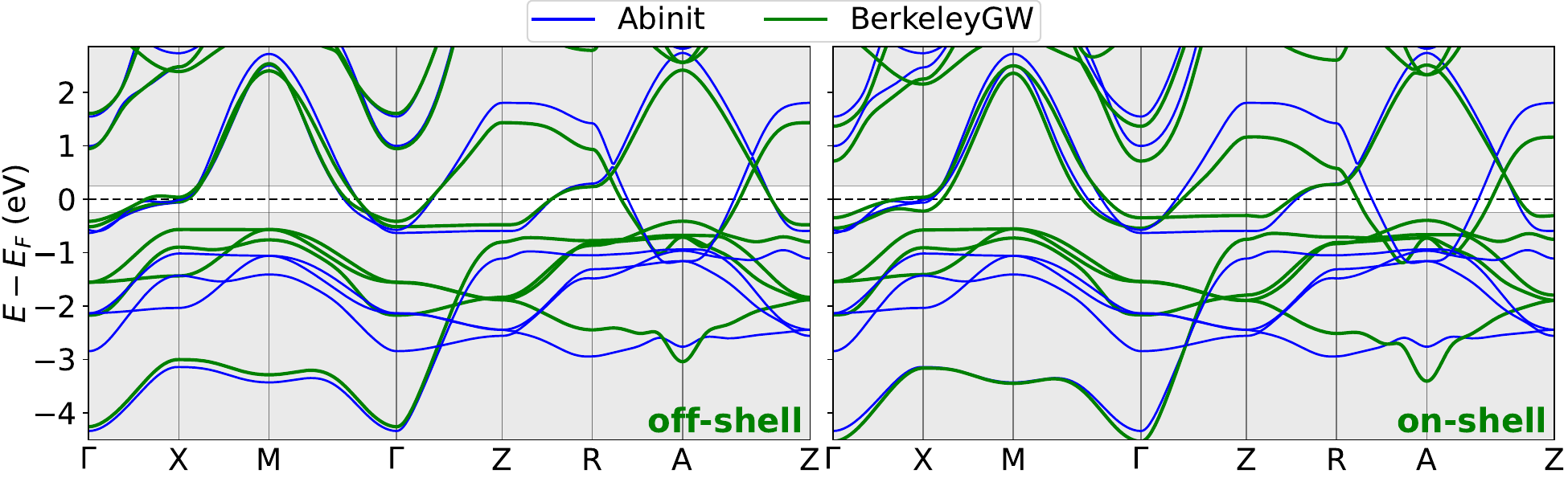} 
\caption{
$GW$ band plots obtained with ABINIT (blue) and BerkeleyGW (green) using the HL plasmon pole model. Although considerably better convergence is obtained with the calculation parameters described in Methods, here we used the same lattice parameters and similar calculation parameters as in \cite{louie22} to ease the comparison. 
The white window indicates $\pm250$ meV around $E_F$. The ABINIT results in (a) and (b) are the same and correspond to the solution of the linearized Dyson equation taking into account the quasiparticle renormalization factor $Z$ (off-shell). The BerkeleyGW results in (a) correspond to the same off-shell solution while in (b) they are obtained on-shell. Specifically, to calculate quasiparticle energies $\epsilon_i$ one should solve the non-linear equation
$
  \epsilon_i = \epsilon_i^\mathrm{DFT} + \langle i|\Sigma(\omega=\epsilon_i) - v_{xc}| i \rangle,
$
where $\epsilon_i$ appears also as argument of the self-energy $\Sigma$. 
The off-shell solution corresponds to the linearized form
$  \epsilon_i = \epsilon_i^\mathrm{DFT} + Z_i \langle i|\Sigma(\omega=\epsilon_i^\mathrm{DFT}) - v_{xc}| i \rangle ,
$
where $Z$ is the quasiparticle renormalization factor defined as
$
Z_i = \left( 1 - \langle i|\partial_\omega \Sigma(\omega=\epsilon_i^\mathrm{DFT})| i \rangle \right)^{-1}.
$
The on-shell solution is a more crude (and therefore not recommended) approximation which consists in neglecting the quasi-particle renormalization factor ($Z=1$):
$ 
 \epsilon_i = \epsilon_i^\mathrm{DFT} 
  + \langle i|\Sigma(\omega=\epsilon_i^\mathrm{DFT}) - v_{xc}| i \rangle .
$
We note that the on-shell procedure in (b) reproduces the band plot reported in \cite{louie22}. 
Further, considering the electronic structures in (a), we find that the corresponding electron-phonon coupling constant $\lambda $ is 0.31 (ABINIT) and 0.36 (BerkeleyGW) while considering the BerkeleyGW one in (b) $\lambda$ becomes 0.41 (and hence explains the larger value reported in \cite{louie22}). With better convergence as described in Methods  $\lambda = 0.163$ is obtained with ABINIT.
}
\label{abinit-vs-berkeleygw}
\end{figure}

\begin{figure}[h!]
    \centering
     \includegraphics[width=.7\textwidth]{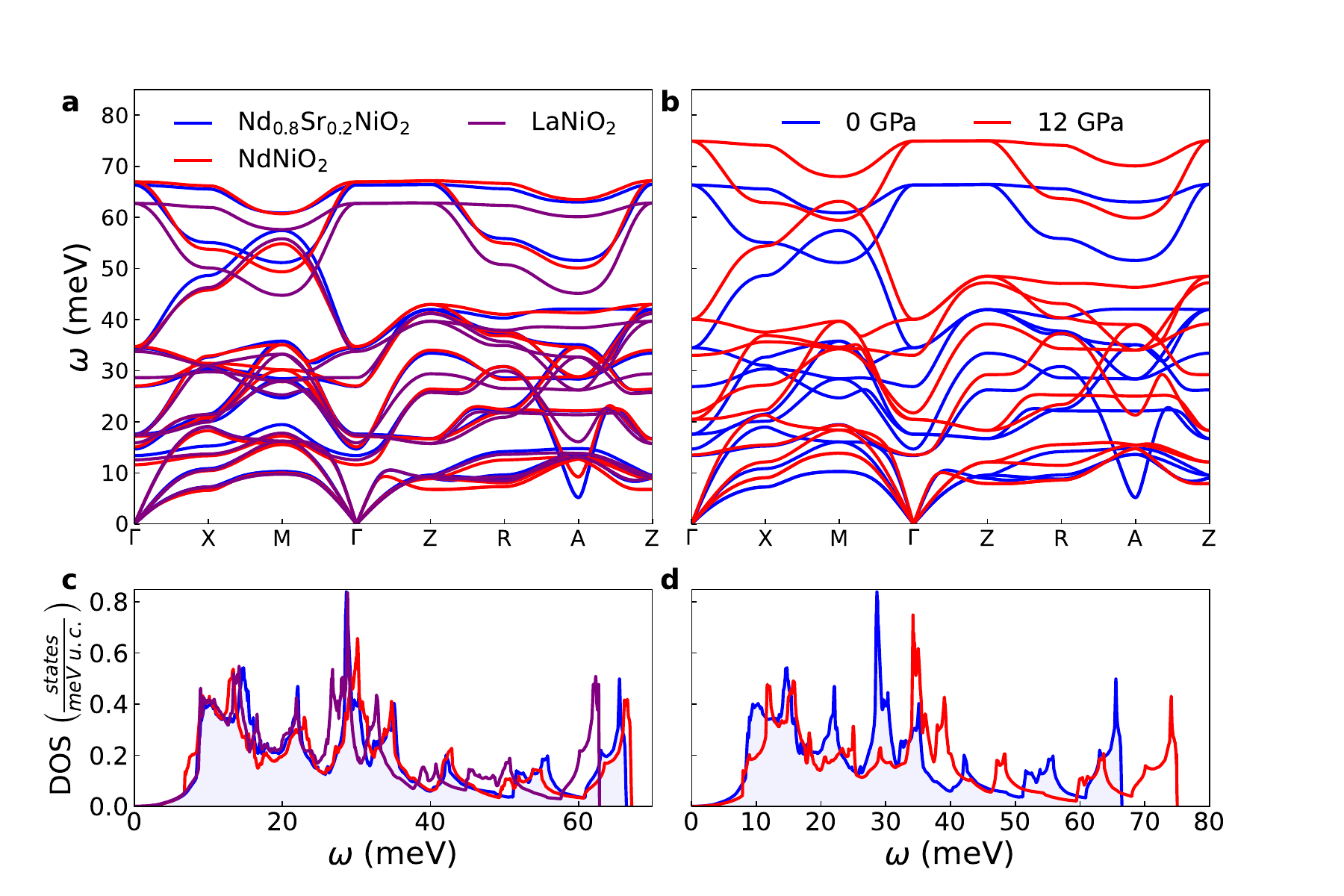}
    \caption{Comparison between (a) the phonon spectrum and (c) the phonon DOS of NdNiO$_2$ (red) and Nd$_{0.8}$Sr$_{0.2}$NiO$_2$ (blue), and comparison between (b) the phonon spectrum and (d) the phonon DOS of Nd$_{0.8}$Sr$_{0.2}$NiO$_2$ at ambient pressure (blue) and at 12 GPa (red). 
    We observe in (c) that the Nd~$\to$~Sr substitution leads to slight softening of the O peaks at 30 and 65 meV while the low-frequency features harden. 
    Pressure, in its turn, produces a general hardening by around 15\% of the overall spectrum (b) and all major features in the DOS (d). }
    \label{phonons-comparison}
\end{figure}

\begin{figure*}[t!]
\includegraphics[width=0.975\textwidth]{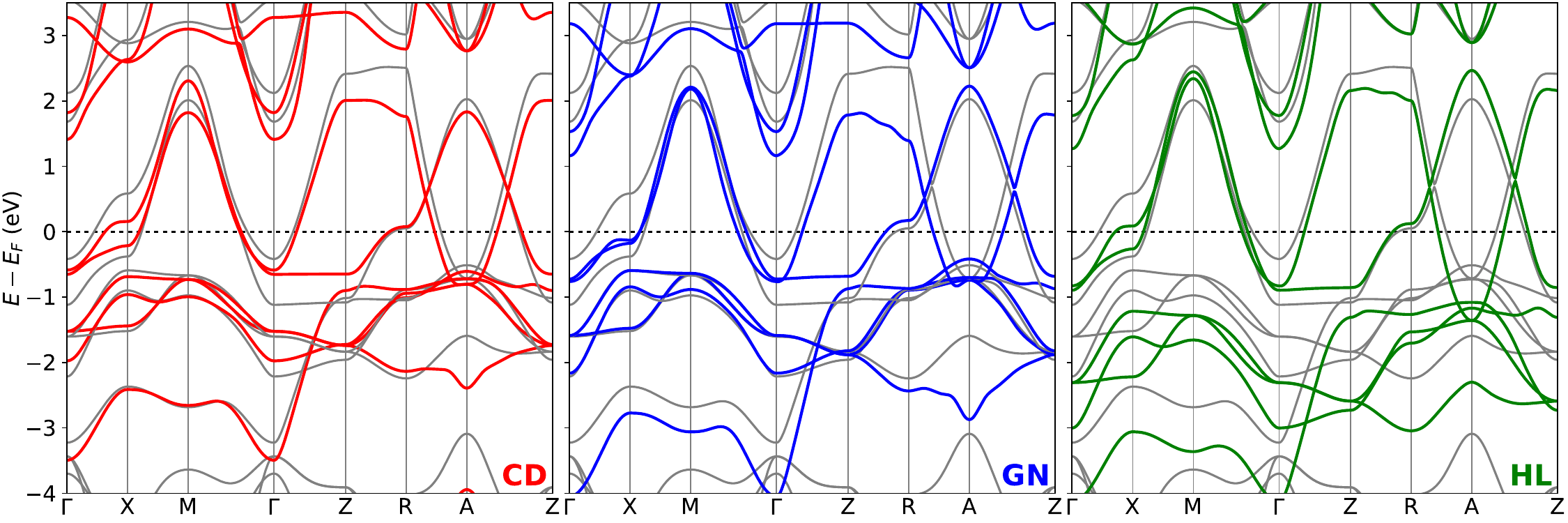}

\vspace{2em}
\includegraphics[width=0.6\textwidth]{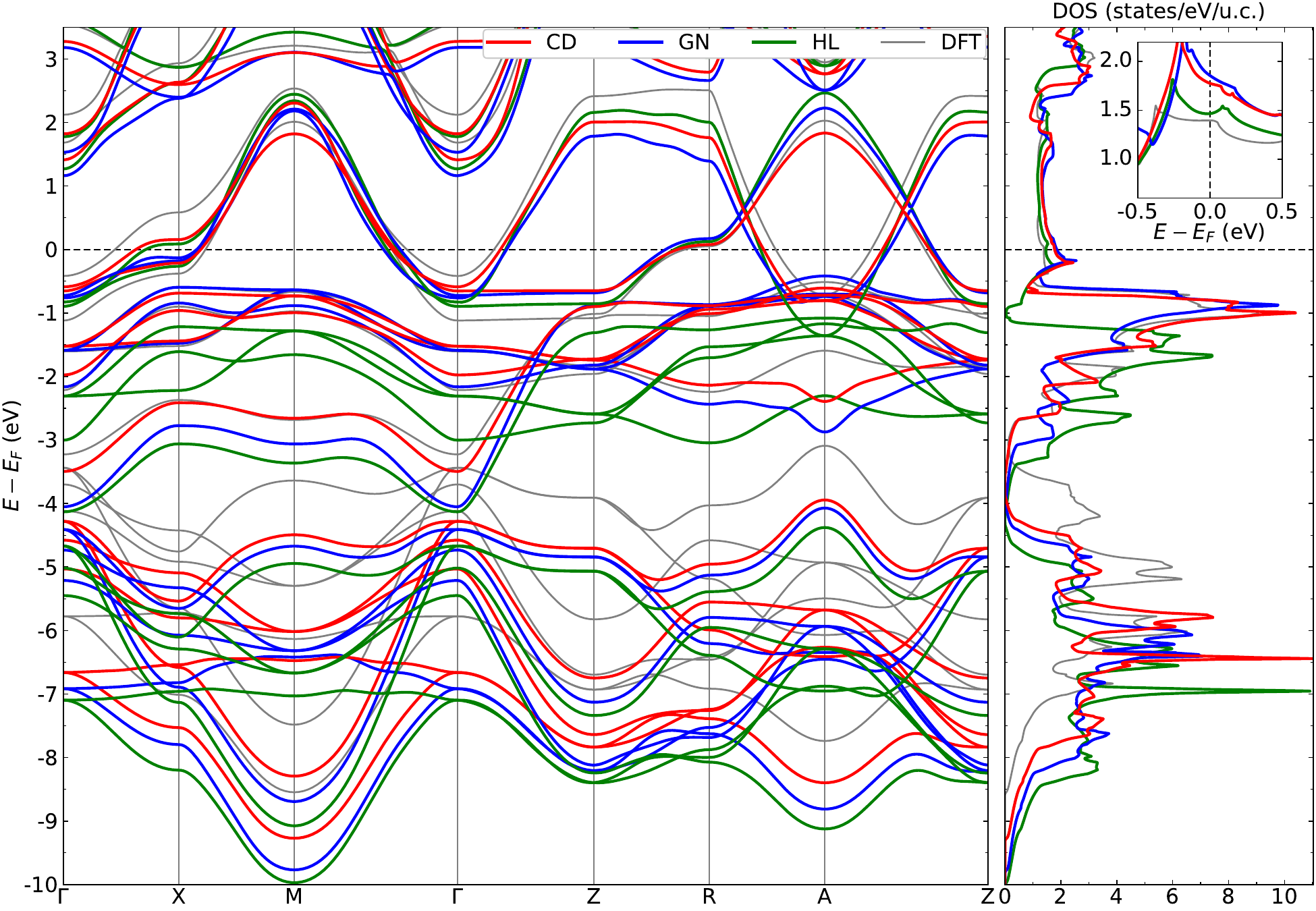}
    \caption{
    Band structure and DOS of NdNiO$_2$ calculated within the $GW$ approximation for the self-energy.  
    The different panels \blue{at the top} correspond to different treatments of the screening via the contour deformation technique (red), the Godby-Needs plasmon-pole model (blue) and the Hybertsen-Louie one (green), with the DFT result in gray. 
    }
    \label{Nd-GW}
\end{figure*}

\begin{figure}
    \centering
    \includegraphics[width=.85\textwidth]{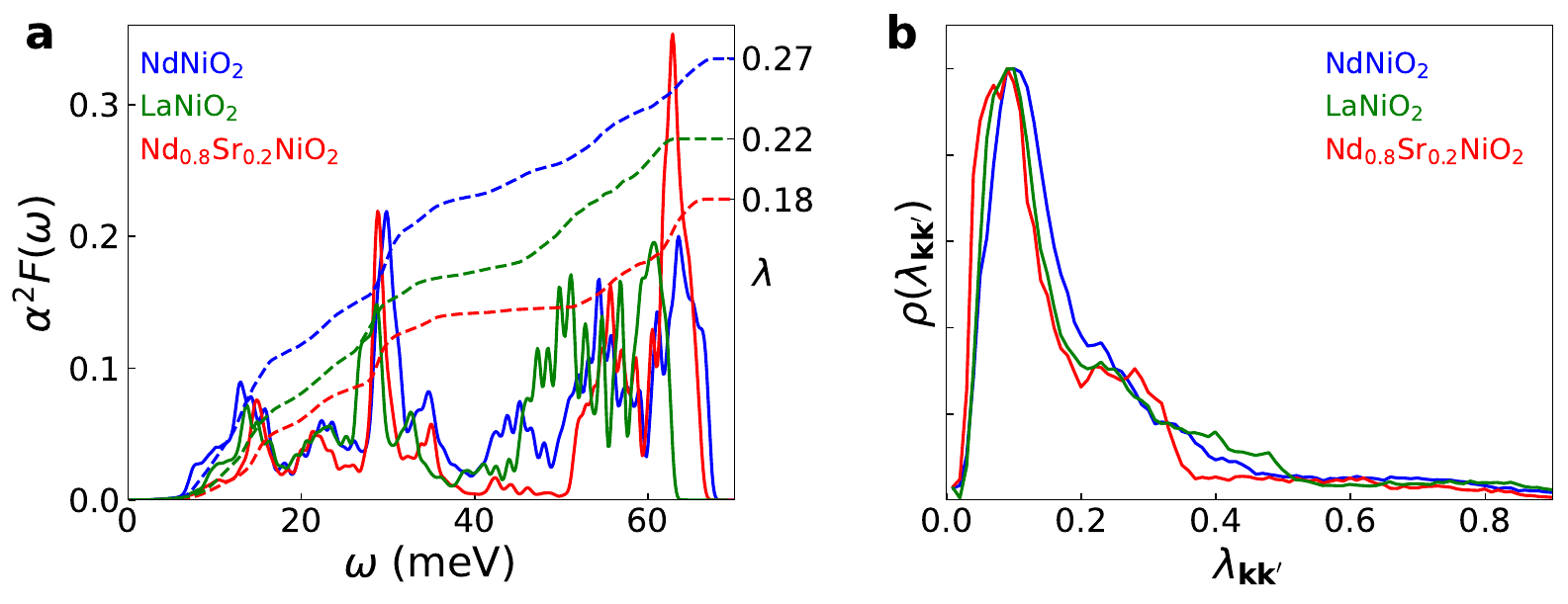}
    \caption{
    Comparison between parent and the hole-doped compounds: (a) Eliashberg function $\alpha^2F(\omega)$ and cumulative coupling constant $\lambda$ and (b) distribution of electron-phonon coupling strength $\lambda_{\mathbf{kk}'}$ calculated from the $GW$-CD. 
    }
    \label{fig:sc_doped_undoped}
\end{figure}

\begin{figure}
    \centering
    \includegraphics[width=.975\textwidth]{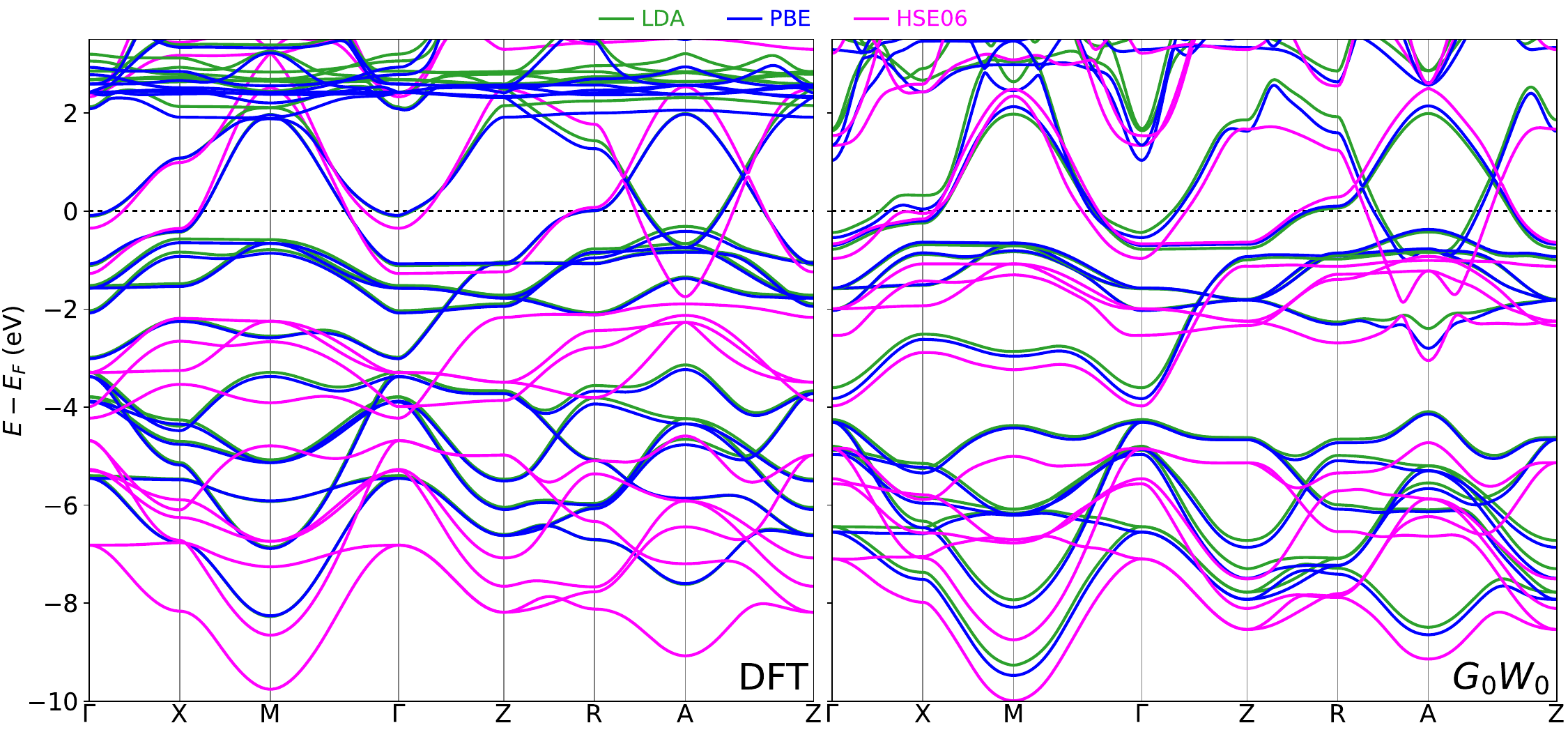}

    \vspace{2em}
    \includegraphics[width=.75\textwidth]{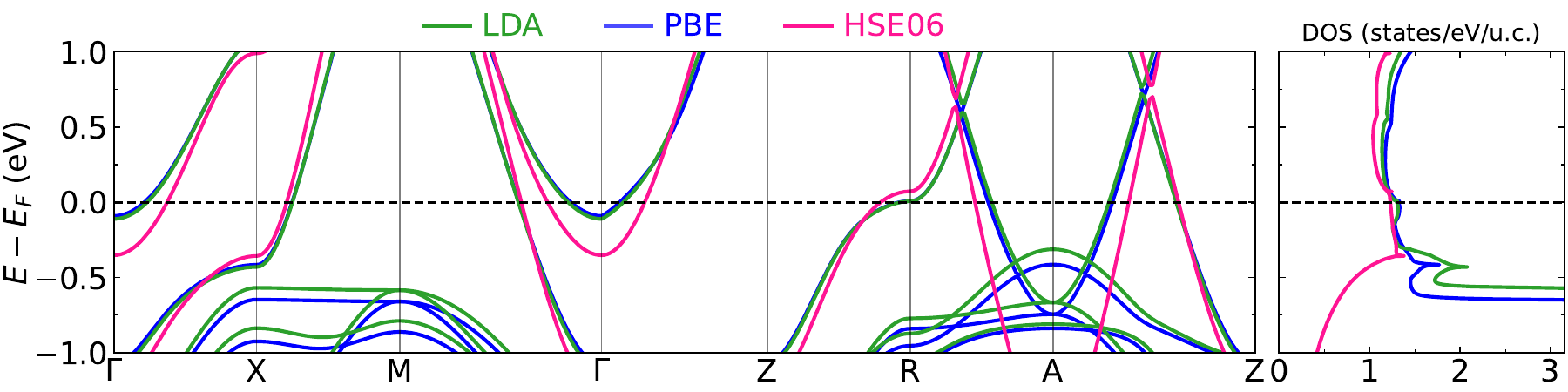}
    \includegraphics[width=.75\textwidth]{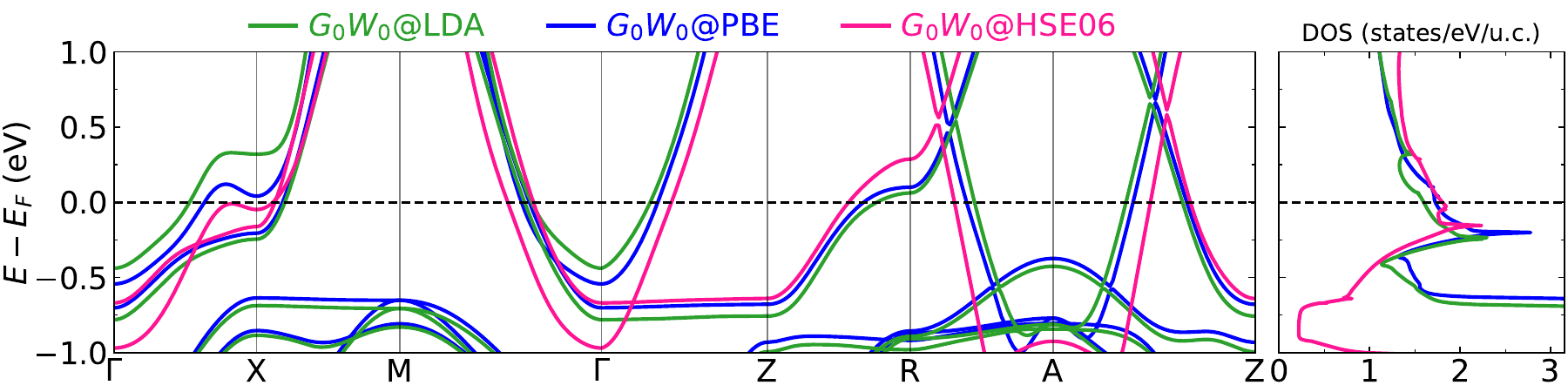}
        \caption{
    \blue{Comparison between the electronic band structure of LaNiO$_2$ computed at the DFT level 
    using LDA, PBE and HSE06 for the exchange-correlation functional and subsequently within the $GW$ approximation using the GN plasmon-pole model. 
    At the DFT level, LDA and PBE yield essentially the same result near the Fermi level with some differences ---mostly related to O-2$p$ and La-4$f$ derived states--- above and below. 
    HSE06, however, introduces important changes everywhere, including the Ni-3$d$ and La-4$d$ bands crossing the Fermi level. 
    Despite these differences, the subsequent $GW$-GN calculation tends to converge towards the same result, especially near the Fermi level (which is the most important region in relation to electron-phonon mediated superconductivity). 
    }}
    \label{fig:starting-dft}
\end{figure}

\begin{figure}
    \centering
    \begin{minipage}[c]{1\textwidth}
    \centering
        \includegraphics[width=0.65\linewidth]{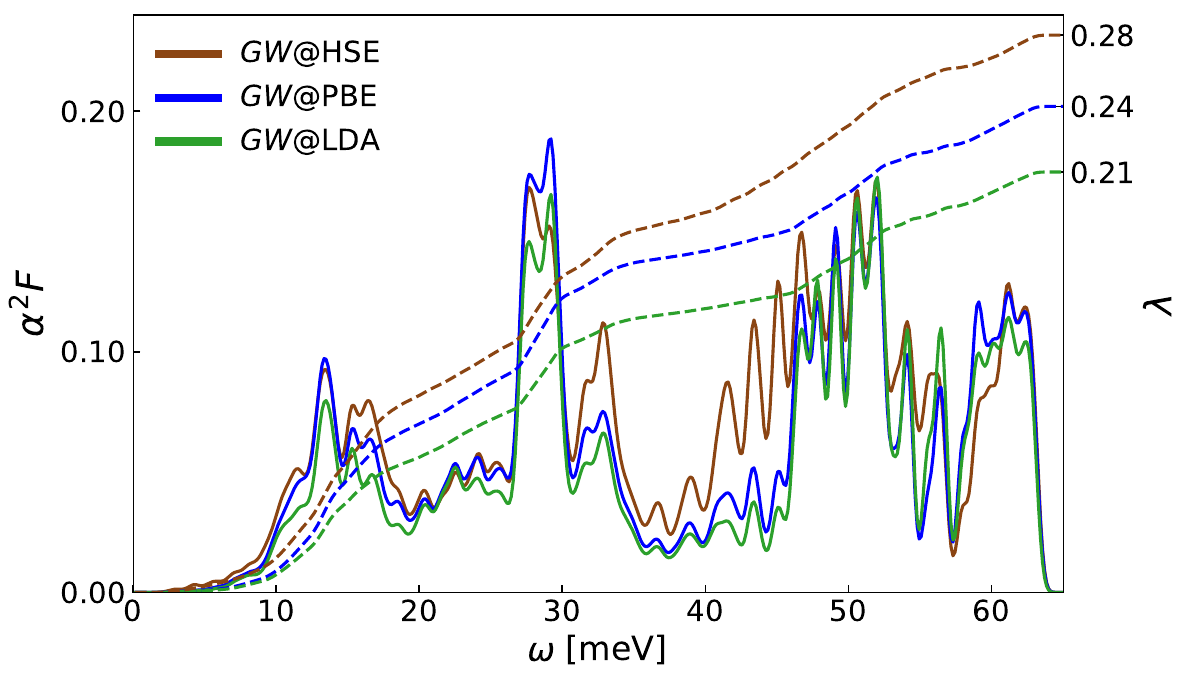}
    \end{minipage}
    \begin{minipage}[c]{0.57\textwidth}
    \centering
        \includegraphics[width=1\linewidth]{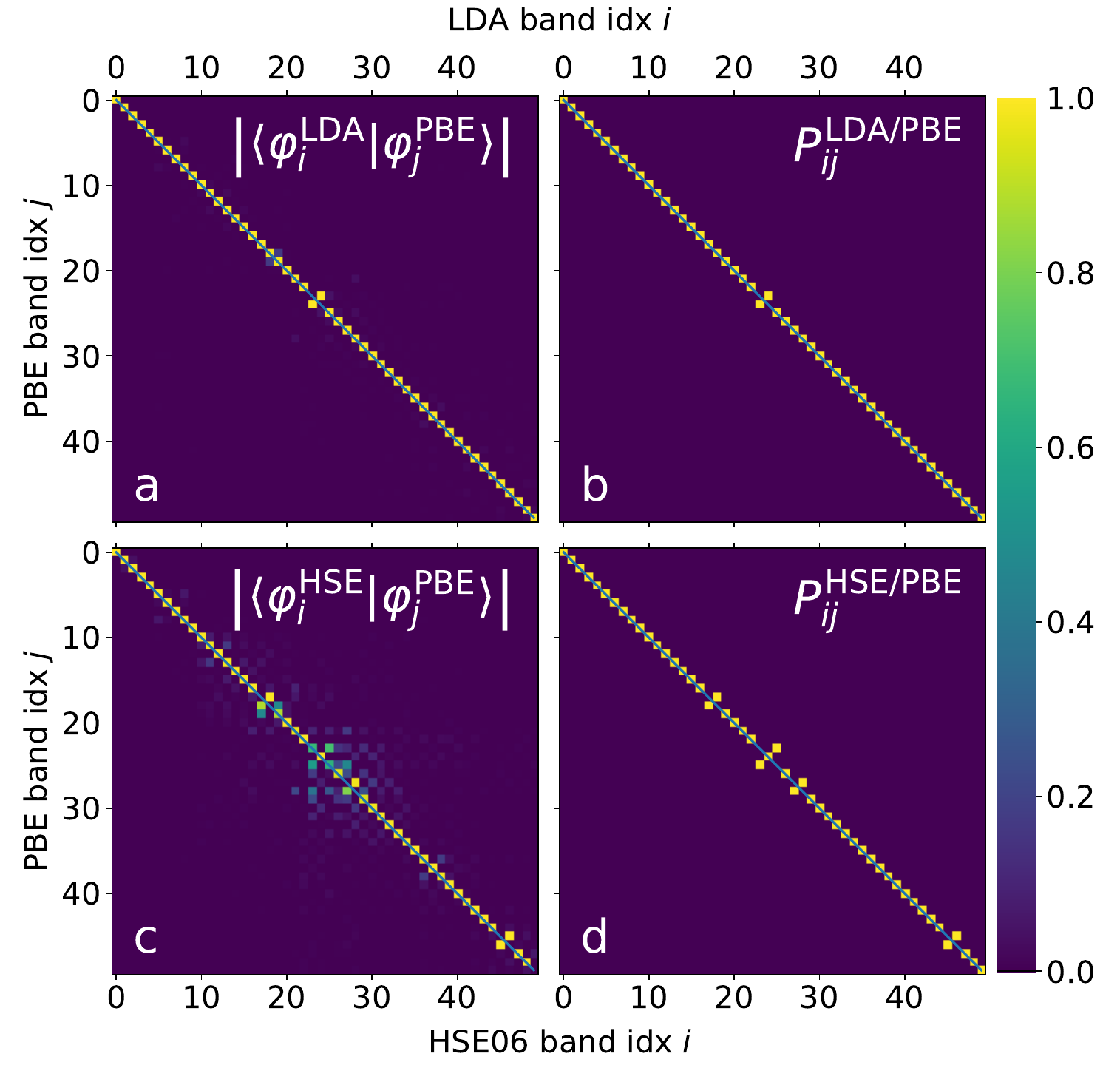}
    \end{minipage}
    \begin{minipage}[c]{0.42\textwidth}
    \centering
        \includegraphics[width=1\linewidth]{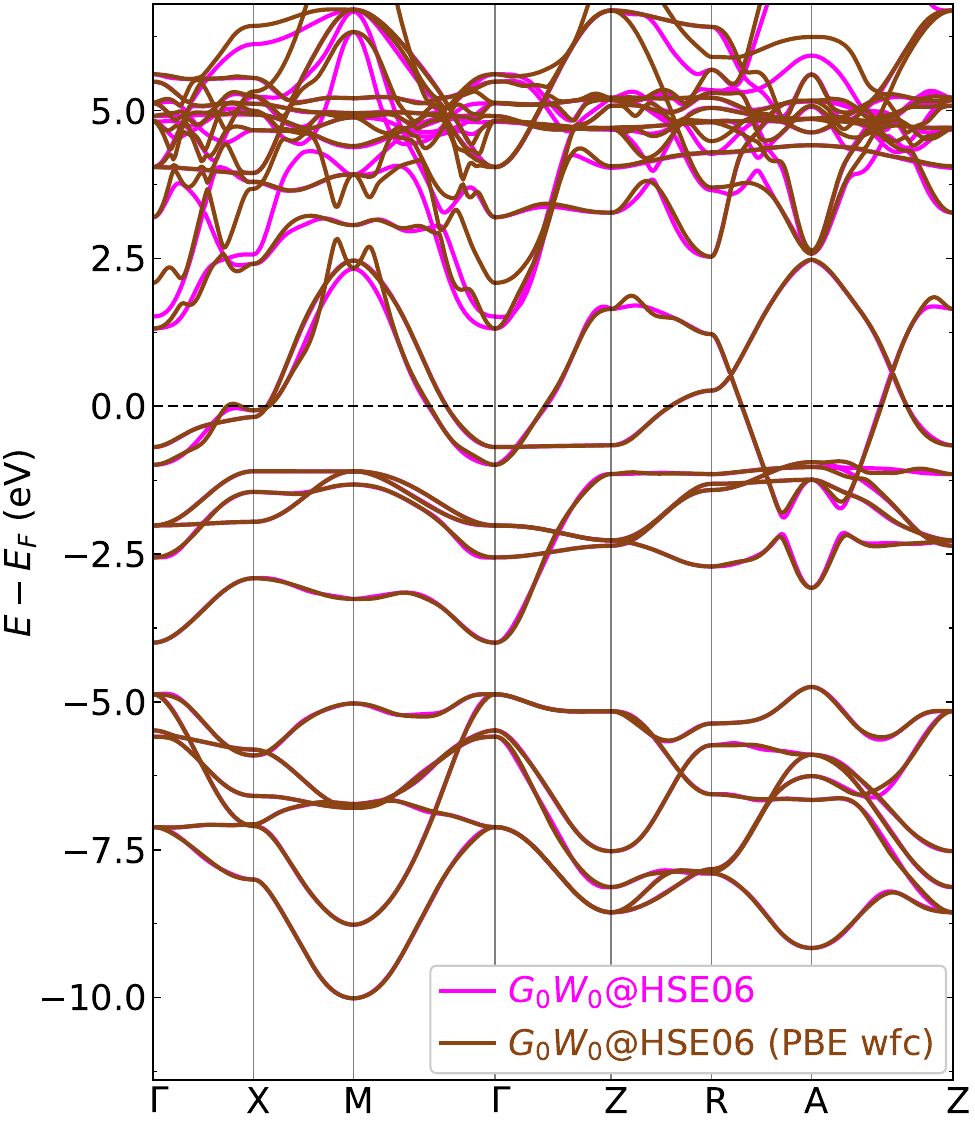}
    \end{minipage}
    \caption{
    \blue{
    (Top panel) Comparison between Eliashberg function $\alpha^2F(\omega)$ and cumulative coupling constant $\lambda$ computed for LaNiO$_2$ from $GW$ calculations using the GN plasmon pole model with different initial electronic structures (LDA, PBE and HSE06).
    In order to single out the effect associated with the renormalization of the quasiparticle energies, the relaxed structure, the phonons and the electron-phonon matrix elements are calculated using PBE in all three cases. In this way, we can also treat the HSE06 functional for which specific phonon calculations would be too expensive.
    In an intermediate step the energies obtained from the different $GW$ calculations (LDA, PBE, and HSE06)  are interpolated on a denser $k$-mesh using a wannierization of PBE wavefunctions. For this procedure to be consistent, the correspondence between energies and their associated wavefunctions must be respected. This point is nontrivial because, for a given $\mathbf k$, the order of energies for different states may swap compared to $GW$@PBE. We corrected for this by analyzing the scalar product between the wavefunctions $|\langle \psi_i ^{\rm X}|  \psi_j ^{\rm PBE}\rangle|$ (where $\rm{X}\!=\!$~LDA, HSE06) and, if needed, reordered the corresponding energies $\varepsilon_i^{GW@{\rm X}}$. 
    The corresponding permutation matrix $P_{ij}$ is obtained as follows: For each $i$, we look for the maximal value of $|\langle \psi_i ^{\rm X}|  \psi_j ^{\rm PBE}\rangle|$. The corresponding $j$ is then taken to define $P_{ij} =1$, while the rest of values are put to 0. 
    (Bottom left) The panels (a) and (b) illustrate $|\langle \psi_i ^{\rm X}|  \psi_j ^{\rm PBE}\rangle|$ and the corresponding permutation matrix for $\bold{k}=(1/3,1/6,1/2)$ in the LDA case. As we see, one state is swapped compared to PBE. (c) and (d) illustrate the same for HSE06. In this case, the region between band indices 21 and 34 where the mixing of the states is the highest corresponds to the 4$f$ bands. We see that for these states PBE and HSE06 wavefunctions do not have a clear one-to-one correspondence, which has a non-negligible but not dramatic influence on the quality of the Wannier interpolation. 
    (Bottom right) Wannier interpolation of $\varepsilon ^{GW@{\rm HSE06}}$ energies using PBE Wannier functions (brown) compared to the ``pure" Wannier interpolation using HSE06 Wannier functions (magenta). Even if the former shows some artifacts above the Fermi level, we obtain a satisfactory fit around the Fermi level which validates {\it a posteriori} the use of the Wannier functions from PBE for the wannier interpolation of LDA/HSE06 eigenvalues.
    \label{fig:lambda-starting}}
    }
\end{figure}

\begin{figure}
    \centering
    \includegraphics[width=.65\linewidth]{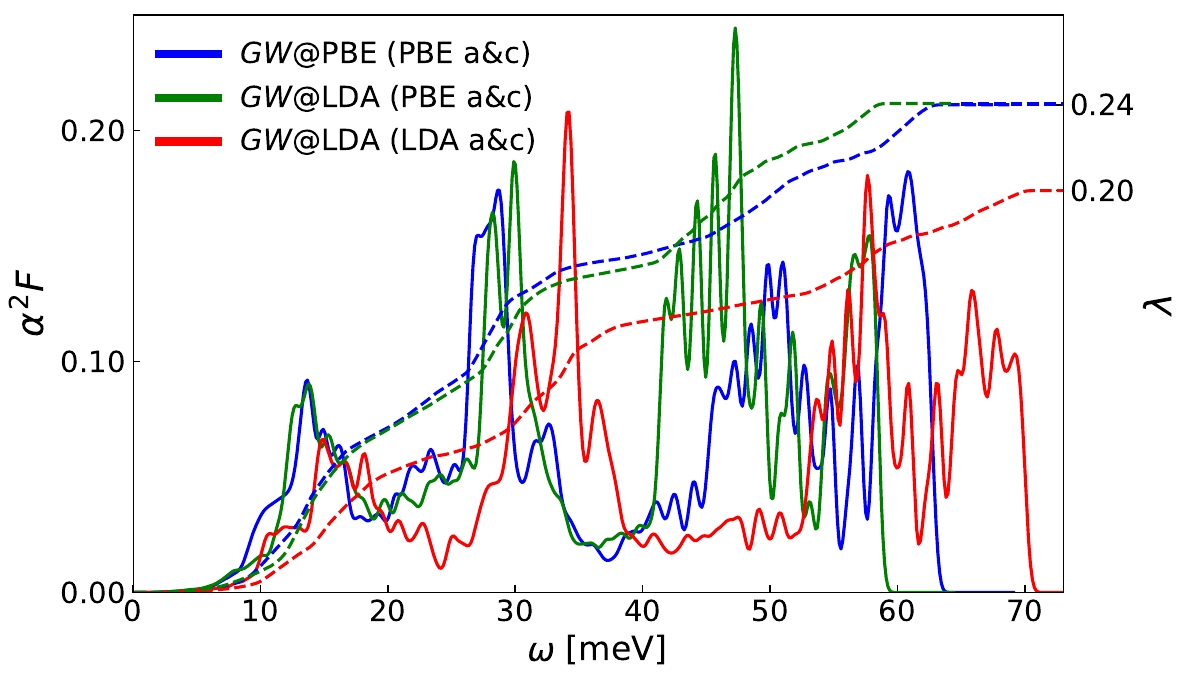}
    \caption{
    \blue{
    Comparison between Eliashberg function $\alpha^2F(\omega)$ and cumulative coupling constant $\lambda$ computed for LaNiO$_2$ from $GW$ calculations using different phonons: LDA with LDA optimized lattice parameters, LDA with PBE optimized lattice parameters and PBE with PBE optimized lattice parameters.  
    }
    }
    \label{fig:lambda-starting-phonons}
\end{figure}

\begin{figure}[h!]
    \centering
    \includegraphics[width=0.495\textwidth]{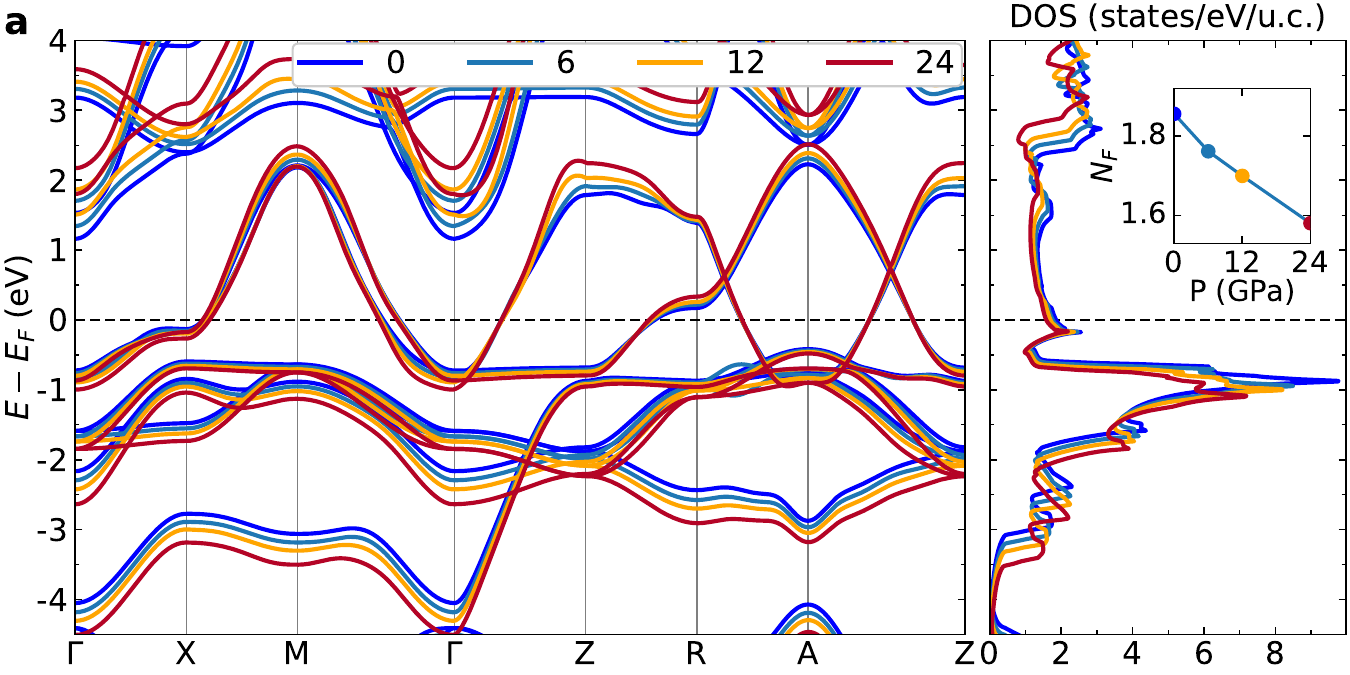}
    \includegraphics[width=0.495\textwidth]{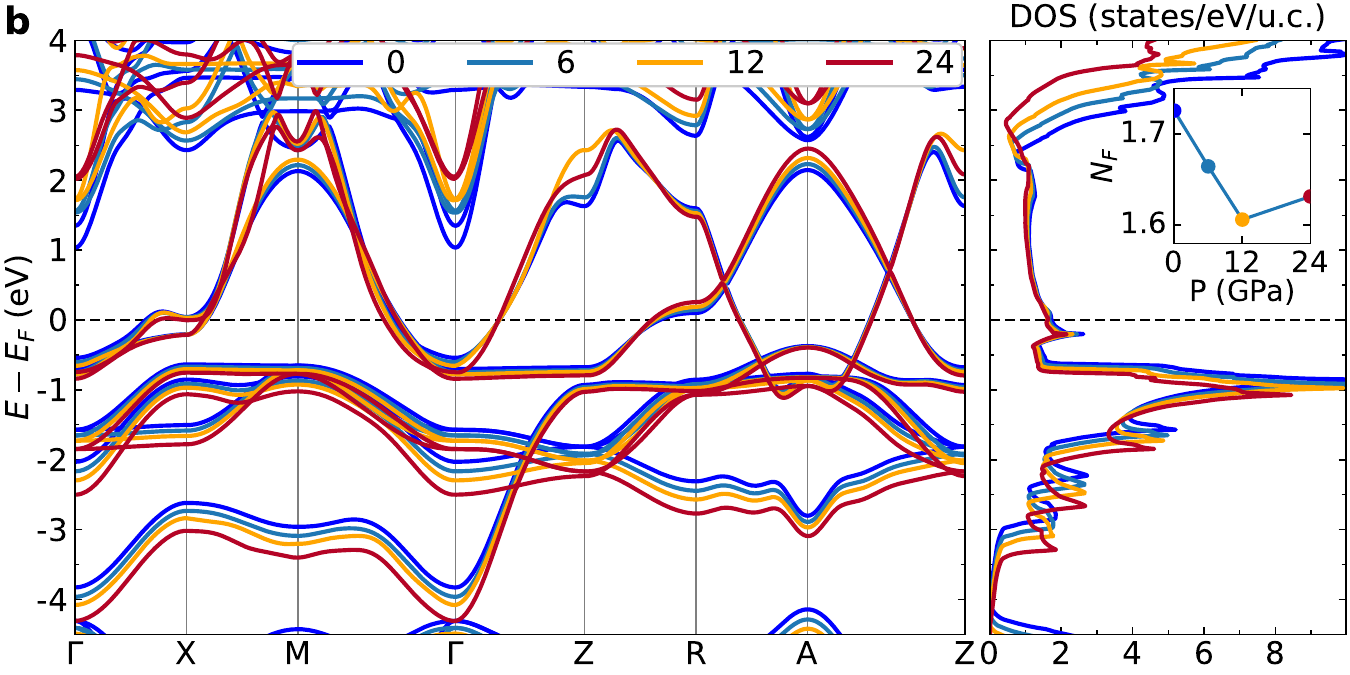}
 
    \caption{Computed $GW$ band structure and DOS of NdNiO$_2$ (left) and LaNiO$_2$ (right) as a function of pressure. The initial self-doping of the Fermi surface is further enhanced at $\Gamma$ with pressure resulting in the swapping of the Ni-3$d_{x^2-y^2}$ and $R$-3$d_{z^2}$ bands along the $\Gamma$-X-M-$\Gamma$ path. At A, in contrast, the self-doping remains essentially the same. In addition, the there is a concomitant increase in the bandwidths that further results into a decrease of the density of states at the Fermi level.
       \label{fig:Nd-GW-bands-pressure}}
\end{figure}

\begin{figure}
    \centering
    \includegraphics[width=.85\textwidth]{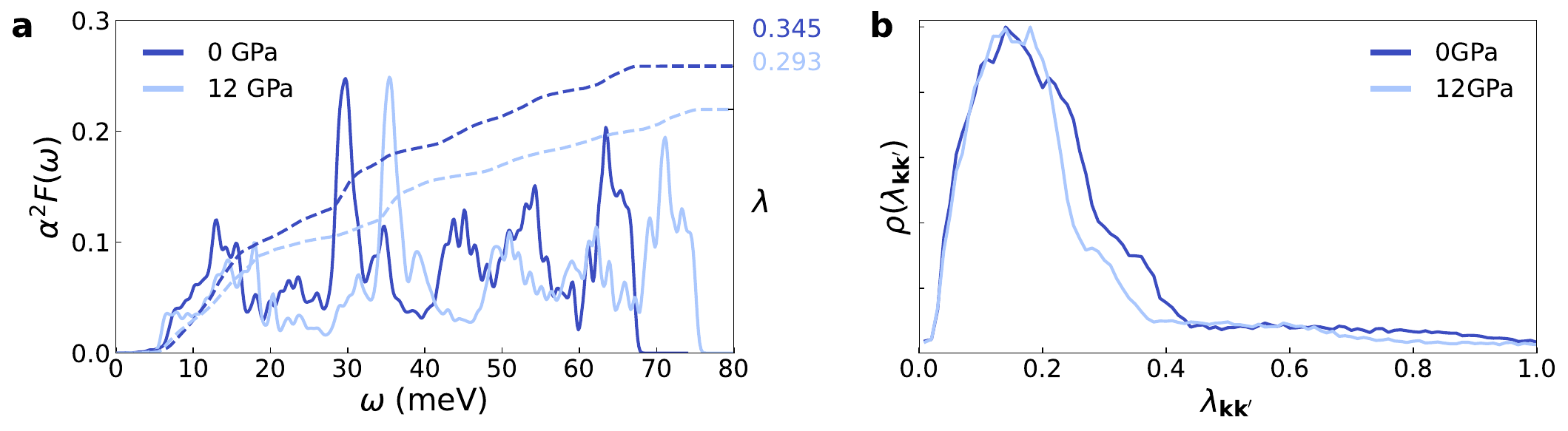}
    \caption{
    (a) calculated Eliashberg function $\alpha^2F(\omega)$ and cumulative coupling constant $\lambda$ and (b) distribution of electron-phonon coupling strength $\lambda_{\mathbf{kk}'}$ of NdNiO$_2$ as a function of pressure.
    }
    \label{fig:nd_pressure}
\end{figure}

\end{document}